# Energy fluctuations and the ensemble equivalence in Tsallis statistics

Liu Liyan, Du Jiulin

*Department of Physics, School of Science, Tianjin University, Tianjin 300072, China*

**Abstract**: We investigate the general property of the energy fluctuation for the canonical ensemble in Tsallis statistics and the ensemble equivalence. By taking the ideal gas and the non-interacting harmonic oscillators as examples, we show that, when the particle number $N$ is large enough, the relative fluctuation of the energy is proportional to $1/N$ in the new statistics, instead of $1/\sqrt{N}$ in Boltzmann-Gibbs statistics. Thus the equivalence between the microcanonical and the canonical ensemble still holds in Tsallis statistics.

**I. Introduction**

In statistical mechanics, there are three different ensembles, namely, the micro-canonical, the canonical and the grand canonical ensemble [1]. In the micro-canonical ensemble, the systems are isolated that both the energy and the particle numbers are fixed. In the canonical ensemble, the systems are closed that the particle number conserves and the energy is a variable, thus, in principle, the energy can take any values from zero to infinity. In the grand canonical ensemble, the systems are open ones that both the energy and the particle numbers are variables. Therefore, the three ensembles seem to be different from each other. However, for the canonical ensemble in Boltzmann-Gibbs (B-G) statistics, we know that, although the energy of a system can take any value between zero and infinity, the relative fluctuation of energy of a macroscopic system is so small that it can be negligible. The energy of a system is almost equal to the mean value, thus being the same as that in the micro-canonical ensemble. The relative fluctuations of energy and particle number in the grand canonical ensemble are also negligible. Therefore, the three ensembles in B-G statistics are equivalent to each other; the result obtained by one ensemble is identical to that by the other two.

In recent years, there has been an increasing focus on a new statistical theory, nonextensive statistics or Tsallis statistics [2]. It is thought that Tsallis statistics is a useful generalization of B-G statistics and is suitable for the statistical description of the nonequilibrium systems with long-rang interactions [3-5 and the references in]. Some applications in these systems are investigated [6-9 and the references in]. Many problems related to the canonical ensemble are discussed in Tsallis statistics [10-14]. As mentioned above, in B-G statistics the relative fluctuation of energy can be negligible and the ensembles are equivalent. Then in Tsallis statistics, a natural question is whether the relative energy fluctuation is still negligible or whether the ensembles are still equivalent. In this paper, we investigate the energy fluctuation in the canonical ensemble in Tsallis statistics and check the equivalence of microcanonical and canonical ensemble.

This paper is organized as follows. In section II, we present a brief review on the canonical distribution using the optimal Lagrange multipliers (OLM) method. In section III, we investigate the energy fluctuation in the canonical ensemble and find the general expressions. In section IV, we take the ideal gas and the harmonic oscillator as examples to study the property of the energy fluctuation. Finally in section V, the summary and conclusion are given.

**II. The canonical distribution in the OLM formalism**

In Tsallis statistics, there are several forms of the probability distribution functions based on the different constraints [15-17]. Here, we adopt the canonical distribution using the so-called OLM method [15]. Now let us give a brief review on this method.

Tsallis entropy [18] is

$$S_q = k \frac{\int (f^q - f) d\Omega}{1-q}, \quad (1)$$

where $k$ is the Boltzmann constant, $f$ is the distribution function, $\Omega$ is the phase space element, and $q$ is the nonextensive parameter. The basic property of Tsallis entropy is the nonadditivity for $q \neq 1$, and for $q \to 1$ it reduces to B-G entropy [18]. In the OLM formalism, the probability distribution function is obtained by recourse to the maximum entropy principle. Namely, one maximizes Tsallis entropy (1) subject to the following constraints [15]

$$\int f d\Omega - 1 = 0 \quad (2)$$

$$\int f^q (H - \langle U \rangle_q) d\Omega = 0. \quad (3)$$

where $H$ is the Hamiltonian of the system, and $\langle U \rangle_q$ is the mean energy which is defined as

$$\langle U \rangle_q \equiv \frac{\int f^q H d\Omega}{\int f^q d\Omega}. \quad (4)$$

Using the Lagrange multipliers $\alpha$ and $\beta$, the extreme solution subject to constraints (2) and (3) can be obtained from

$$\delta S_q - \alpha \left( \int f d\Omega - 1 \right) - \beta \left[ \int f^q (H - \langle U \rangle_q) d\Omega = 0 \right]. \quad (5)$$

After several calculations one can get the canonical distribution function

$$f = \frac{\left[ 1 - (1-q)\beta(H - \langle U \rangle_q) \right]^{\frac{1}{1-q}}}{Z_q}, \quad (6)$$

where $\beta = 1/kT$, and $Z_q$ is the partition function given by

$$Z_q = \int \left[ 1 - (1-q)\beta(U - \langle U \rangle_q) \right]^{\frac{1}{1-q}} d\Omega. \quad (7)$$

Normalizing the distribution function (6) leads to a relation

$$\int f^q d\Omega = Z_q^{1-q}. \quad (8)$$

Furthermore, from eqs. (6) and (7) it can be proved that

$$Z_q = \int \left[ 1 - (1-q)\beta(U - \langle U \rangle_q) \right]^{\frac{q}{1-q}} d\Omega, \quad (9)$$

**III. The energy fluctuations in Tsallis statistics**

Here we will discuss the energy fluctuation for the canonical ensemble by employing the distribution function in the OLM formalism. From the probability distribution function eq. (6), we can obtain

$$f^{1-q} = \frac{1-(1-q)\beta(H-\langle U\rangle_q)}{Z_q^{1-q}}. \tag{10}$$

After multiplying eq. (10) by $Hf^q$ and integrating it over the phase space, we get

$$\int Hfd\Omega = \int \frac{f^q H\left[1-(1-q)\beta(H-\langle U\rangle_q)\right]}{Z_q^{1-q}} d\Omega. \tag{11}$$

With the help of eq. (8), we finds

$$\int Hfd\Omega = \frac{\int f^q H\left[1-(1-q)\beta(H-\langle U\rangle_q)\right]d\Omega}{\int f^q d\Omega}$$

$$= \left\langle H\left[1-(1-q)\beta(H-\langle U\rangle_q)\right]\right\rangle_q, \tag{12}$$

$$= \langle U\rangle_q - (1-q)\beta\left(\langle U^2\rangle_q - \langle U\rangle_q^2\right)$$

where the definition of the expectation value of an arbitrary physical quantity $F$, $\langle F\rangle_q \equiv \int f^q F d\Omega / (\int f^q d\Omega)$, in the OLM formalism is used. Then performing the differentiation $\partial/\partial\beta$ on the left hand side of eq. (13), we derive

$$\frac{\partial(\int Hfd\Omega)}{\partial\beta} = \frac{\int H\left(-(H-\langle U\rangle_q)+\beta\partial\langle U\rangle_q/\partial\beta\right)\left[1-(1-q)\beta(H-\langle U\rangle_q)\right]^{\frac{q}{1-q}}d\Omega}{Z_q}$$

$$-\int Hfd\Omega \frac{\int\left(-(H-\langle U\rangle_q)+\beta\partial\langle U\rangle_q/\partial\beta\right)\left[1-(1-q)\beta(H-\langle U\rangle_q)\right]^{\frac{q}{1-q}}d\Omega}{Z_q}. \tag{13}$$

Using eq. (9), we obtain

$$\frac{\partial(\int Hfd\Omega)}{\partial\beta} = -\left(\langle U^2\rangle_q - \langle U\rangle_q^2\right) + \beta\langle U\rangle_q \frac{\partial\langle U\rangle_q}{\partial\beta} - \beta\frac{\partial\langle U\rangle_q}{\partial\beta}\int Hfd\Omega. \tag{14}$$

Insertion of eq. (12) into (14) yields

$$\frac{\partial\langle U\rangle_q}{\partial\beta} + q\left(\langle U^2\rangle_q - \langle U\rangle_q^2\right) - (1-q)\beta\frac{\partial\left(\langle U^2\rangle_q - \langle U\rangle_q^2\right)}{\partial\beta} = (1-q)\beta^2\frac{\partial\langle U\rangle_q}{\partial\beta}\left(\langle U^2\rangle_q - \langle U\rangle_q^2\right), \tag{15}$$

or it can be written as

$$(1-q)\beta\frac{\partial\langle(\Delta U)^2\rangle_q}{\partial\beta} + \left[(1-q)\beta^2\frac{\partial\langle U\rangle_q}{\partial\beta} - q\right]\langle(\Delta U)^2\rangle_q = \frac{\partial\langle U\rangle_q}{\partial\beta}, \tag{16}$$

where $\langle(\Delta U)^2\rangle_q \equiv \left\langle\left(U-\langle U\rangle_q\right)^2\right\rangle_q = \langle U^2\rangle_q - \langle U\rangle_q^2$ is the energy fluctuation in Tsallis statistics. Obviously, eq. (16) is a first-order differential equation for the energy fluctuation. In the limit of $q\to 1$, it reduces to the result in B-G statistics,

$$\langle(\Delta U)^2\rangle = -\frac{\partial\langle U\rangle_q}{\partial\beta} = \frac{C_V}{k\beta^2}, \tag{17}$$

where $C_V = \partial\langle U\rangle_q/\partial T$ is the heat capacity. The solution of eq.(16) is

$$\langle(\Delta U)^2\rangle_q = C\beta^{\frac{C_V}{k}+\frac{q}{1-q}} + \frac{C_V}{(1-q)C_V+(2-q)k}\frac{1}{\beta^2}, \tag{18}$$

where $C$ is the integration constant. As one knows, in the limit $\beta \to \infty$, or equivalently, when the temperature approaches zero, all the particles occupy the ground state and the energy fluctuation should be zero. But, the first term, $\beta^{C_V/k+q/(1-q)}$, is infinity. So the integration constant should be taken zero to make the term vanish. The final result of the energy fluctuation is

$$\langle (\Delta U)^2 \rangle_q = \frac{C_V}{(1-q)C_V + (2-q)k} \frac{1}{\beta^2}. \tag{19}$$

where the nonextensive parameter should be $0 < q < 1$ for the positivity. According to the relative fluctuation of energy, $\sqrt{\langle (\Delta U)^2 \rangle_q}/\langle U \rangle_q$, we have

$$\frac{\sqrt{\langle (\Delta U)^2 \rangle_q}}{\langle U \rangle_q} = \frac{1}{\beta \langle U \rangle_q} \sqrt{\frac{C_V}{(1-q)C_V + (2-q)k}}. \tag{20}$$

When the limit $q \to 1$, the standard result $\sqrt{\langle (\Delta U)^2 \rangle}/\langle U \rangle = \sqrt{C_V/k}/(\beta \langle U \rangle)$ in B-G statistics is recovered. Thus the general expressions of the energy fluctuation are obtained in Tsallis statistics and the forms in B-G statistics are generalized.

**IV. The Applications**

Now we take the ideal gas and the harmonic oscillator system as examples to apply the above formula to analyze their energy fluctuations.

1. *The ideal gas*

The energy of the ideal gas in terms of OLM formalism [19] is $\langle U \rangle_q = 3N/(2\beta)$, which is an extensive quantity independent of the nonextensive parameter $q$. The heat capacity is $C_V = 3Nk/2$. From eq. (19) and (20), the energy fluctuation reads

$$\langle (\Delta U)^2 \rangle_q = \frac{3N}{(1-q)3N + 4 - 2q} \frac{1}{\beta^2}, \tag{21}$$

which is nonextensive quantity, dependent on the parameter $q$. The relative fluctuation of energy is

$$\frac{\sqrt{\langle (\Delta U)^2 \rangle_q}}{\langle U \rangle_q} = \frac{2}{3N} \sqrt{\frac{1}{1-q+(4-2q)/(3N)}}, \tag{22}$$

The forms in B-G statistics, $\langle (\Delta U)^2 \rangle_q = 3N/(2\beta^2)$, and $\sqrt{\langle (\Delta U)^2 \rangle}/\langle U \rangle = \sqrt{2/3N}$, are recovered in the limit $q \to 1$, respectively.

If the value of $q$ is not in the vicinity of unity, from eq. (22) we find that, when $N \gg (4-2q)/[3(1-q)]$, the relative fluctuation of energy can be written approximately as

$$\frac{\langle (\Delta U)^2 \rangle_q}{\langle U \rangle_q} \approx \frac{2}{3\sqrt{1-q}} \frac{1}{N}, \tag{23}$$

being proportional to $1/N$. As compared with the result, the relative fluctuation of energy is proportional to $1/\sqrt{N}$, in B-G statistics, we get the relative width of the canonical energy distribution is much smaller than that in Tsallis statistics. Obviously, the relative fluctuation of energy, eq. (23), is quite negligible when the particle number is large. Thus the equivalence between the microcanonical and the canonical ensemble still holds in Tsallis statistics.

2. *The noninteracting harmonic oscillators*

The Hamiltonian of the system with $N$ non-interacting harmonic oscillators can be written as

$$H = \sum_{i=1}^{N} \frac{p_i^2}{2m} + \frac{1}{2} m\omega^2 q_i^2, \qquad (24)$$

where $p_i$ and $q_i$ are the momentum and the coordinate of $i$th oscillator, respectively; $m$ is its mass, and $\omega$ is its frequency. From eq. (6) the distribution function is

$$f = \frac{\left[1-(1-q)\beta(H-\langle U\rangle_q)\right]^{\frac{1}{1-q}}}{Z_q}, \qquad (25)$$

with the partition function $Z_q$ given by

$$Z_q = \int \left[1-(1-q)\beta(H-\langle u\rangle_q)\right]^{\frac{1}{1-q}} d\Omega, \qquad (26)$$

where $d\Omega = \left(1/(N!h^N)\right)\prod_{i=1}^{N} dp_i dq_i$ and $h$ is the Planck's constant. According to eq. (4), we obtain the mean energy

$$\langle U\rangle_q = N/\beta, \qquad (27)$$

which is identical to the result in B-G statistics. Then the energy fluctuation (19) of the $N$-oscillator system reads

$$\langle (\Delta U)^2 \rangle_q = \frac{N}{(1-q)N+2-q} \frac{1}{\beta^2}. \qquad (28)$$

The relative fluctuation of energy (20) is

$$\frac{\sqrt{\langle (\Delta U)^2 \rangle_q}}{\langle U \rangle_q} = \frac{1}{N}\sqrt{\frac{1}{(1-q)+(2-q)/N}}. \qquad (29)$$

It is easy to verify that in the limit $q \to 1$, eqs. (28) and (29) reduce to the standard forms in B-G statistics, $\langle (\Delta U)^2 \rangle_q = N/\beta^2$, $\sqrt{\langle (\Delta U)^2 \rangle}/\langle U \rangle = \sqrt{1/N}$. Just as the behavior for the ideal gas in the case of $q \neq 1$, when $N \square (2-q)/(1-q)$, the relative fluctuation of energy for the harmonic oscillators is also proportional to $1/N$. Therefore, for the macroscopic system where the particle numbers are assumed to be infinity, the equivalence between the microcanonical and the canonical ensemble is reproduced in Tsallis statistics.

**V. Summary and conclusion**

We have investigated the general property of the energy fluctuation for the canonical ensemble in Tsallis statistics and the ensemble equivalence. We have obtained the general expressions for the energy fluctuation and the relative fluctuation of energy. When they are applied to the ideal gas and the non-interacting harmonic oscillators, it is shown that although the energies are extensive and independent on $q$, the energy fluctuations are nonextensive and are dependent on the nonextensive parameter $q$. Furthermore, when the particle number $N$ is large enough, the relative fluctuation of the energy is proportional to $1/N$, instead of to $1/\sqrt{N}$ as that in B-G statistics. Thus, the relative width of the canonical energy distribution is much smaller than that in Tsallis statistics, quite negligible for the large particle number. Therefore, the equivalence between the maicriocanonical and the canonical ensemble still holds in Tsallis statistics.

**Acknowledgements**

We would like to thank the National Natural Science Foundation of China under the grant No.10675088.